# Ultra-thin Topological Insulator Bi$_2$Se$_3$ Nanoribbons Exfoliated by Atomic Force Microscopy

Seung Sae Hong[1], Worasom Kundhikanjana[1], Judy J. Cha[2], Keji Lai[1], Desheng Kong[2], Stefan Meister[2], Michael A. Kelly[2], Zhi-Xun Shen[1], Yi Cui[2,*]

[1]Department of Applied Physics and [2]Department of Materials Science and Engineering, Stanford University, Stanford, CA 94305

*To whom correspondence should be addressed. E-mail: yicui@stanford.edu.

Ultra-thin topological insulator nanostructures, in which coupling between top and bottom surface states takes place, are of great intellectual and practical importance. Due to the weak Van der Waals interaction between adjacent quintuple layers (QLs), the layered bismuth selenide (Bi$_2$Se$_3$), a single Dirac-cone topological insulator with a large bulk gap, can be exfoliated down to a few QLs. In this paper, we report the first controlled mechanical exfoliation of Bi$_2$Se$_3$ nanoribbons (> 50 QLs) by an atomic force microscope (AFM) tip down to a single QL. Microwave impedance microscopy is employed to map out the local conductivity of such ultra-thin nanoribbons, showing drastic difference in sheet resistance between 1~2 QLs and 4~5 QLs. Transport measurement carried out on an exfoliated (≤5 QLs) Bi$_2$Se$_3$ device shows non-metallic temperature dependence of resistance, in sharp contrast to the metallic behavior seen in thick (>50 QLs) ribbons. These AFM-exfoliated thin nanoribbons afford interesting candidates for studying the transition from quantum spin Hall surface to edge states.

The metallic surface states of 3D topological insulators[1-7] are protected from disorder effects such as crystal defects and non-magnetic impurities, promising realization of dissipationless electron transport in the absence of high magnetic fields[8]. After the initial discovery of the 2D quantum spin Hall effect (QSHE) in HgTe quantum wells[7,9], three binary compounds – $Bi_2Se_3$, $Bi_2Te_3$, and $Sb_2Te_3$ were predicted and later confirmed by angle-resolved photoemission spectroscopy (ARPES) as 3D topological insulators[10-13]. In particular, $Bi_2Se_3$ has been studied due to the relatively large bulk band gap (~0.3 eV) and the simple band structure near the Dirac point.[12-17] Many exotic physical phenomena are predicted to emerge in low dimensional nanostructures of $Bi_2Se_3$.[18,19] For example, ultra-thin $Bi_2Se_3$ down to a few (≤5) nanometers is expected to exhibit topologically non-trivial edge states, which serves as a new platform for the 2D QSHE[18]. In addition, tuning of the chemical potential becomes easier than thick $Bi_2Se_3$ due to the suppression of bulk contribution. Fortunately, such ultra-thin $Bi_2Se_3$ can be naturally obtained due to its layered rhombohedral crystal structure; two Bi and three Se atomic sheets are covalently bonded to form one quintuple layer (QL, ~1 nm thick) (Figure 1b), where adjacent QLs are coupled by relatively weak van der Waals interaction. Such anisotropic bonding structure implies that similar to the case of graphene,[20] low dimensional crystals of $Bi_2Se_3$ can be generated by mechanical exfoliation, which has been achieved by several groups[21-23]. However, the obtained flakes are usually irregular in shape and the yield of obtaining ultra-thin flakes is low: the typical reported flakes (~10 nm) are still thick compared with the 2D limit where strong coupling between top and bottom surfaces occurs. Therefore, thinning $Bi_2Se_3$ in a controlled manner is desirable. Here we demonstrate a new method to exfoliate $Bi_2Se_3$ by dragging atomic force microscopy (AFM) tip[24] horizontally across the $Bi_2Se_3$ nanoribbons[14,25-28]. If the horizontal force applied by the AFM tip is large enough to break the Bi-Se bonds, the upper portion of the ribbon will be scratched away. By controlling the lateral force and vertical position of the tip, it is possible to remove most of upper QLs and have intact ultra-thin $Bi_2Se_3$ at the bottom.

Figure 2 illustrates the AFM exfoliation procedure of a $Bi_2Se_3$ nanoribbon. $Bi_2Se_3$ nanoribbons are initially synthesized by vapor-liquid-solid (VLS) method using Au catalyst particles on silicon substrate.[25] The thickness of $Bi_2Se_3$ nanoribbons are 25~100nm in thickness. They are transferred to an oxidized silicon (Si/300nm $SiO_2$) substrate for exfoliation. In order to increase surface friction, the substrate is pre-treated with organic polymer (i.e. Poly methyl methacrylate, PMMA) and annealed (200-400℃) in Argon atmosphere after the nanoribbon transfer. A commercial AFM tip made of silicon is used in contact mode, to exfoliate $Bi_2Se_3$ nanoribbon. The AFM tip approaches a $Bi_2Se_3$ nanoribbon in the horizontal direction, and scans across the nanoribbons (50-100nm thick) at a constant height (Figure 2a). When the tip apex is lower than the top surface of the $Bi_2Se_3$ nanoribbon, horizontal movement applies force and breaks the nanoribbon, leaving some residual intact layers underneath the broken part (Figure 2b - 2e). By repeating this scan on a single nanoribbon multiple times, we are able to thin down thick nanoribbons (~ 100 nm) down to a few QLs, while maintaining the original lateral length about several μm.

After the AFM exfoliation, the topography of the nanoribbons is characterized by the usual AFM function. The observed thicknesses vary from ribbons to ribbons, but usually between 1 nm and 15 nm. A few examples of ultra-thin $Bi_2Se_3$ nanoribbons are shown in Figure 3. Majority of the exfoliated ribbons have uniform thickness (Figure 3a-3d), but sometimes we also observe discrete thickness steps within a single ribbon (Figure 3e, 3f). We note that the thickness of thin nanoribbons changes by multiples of the height of one QL (~1 nm), which is expected from the weak bonding between two adjacent QLs.

The AFM exfoliation mechanism can be understood by comparing the horizontal force due to torsion of the AFM cantilever with the bonding strength of $Bi_2Se_3$. When an AFM tip scans through a nanoribbon while maintaining its height, it is tilted (with respect to x-axis) to go over the sidewall. The torsional rotation of the cantilever produces shear stress, which is linearly proportional to the rotation angle φ and applies torque on the side of the nanoribbon. The maximum force $F_{max}$ applicable by the AFM tip can be estimated from the maximum rotation $\varphi_{max}$ due to the nanoribbon thickness $d$, as the equation below[29],

$$F_{max} = \frac{JG}{LR}\varphi_{max} = \frac{JG}{LR}\sqrt{\frac{d}{R}} \qquad (1)$$

where $J$ is the cantilever's polar moment of inertia, $G$ the shear modulus, $L$ the cantilever length, and $R$ the tip height. In case of a typical AFM tip made of Si ($G$ = 79.9 GPa) on 100 nm thick nanoribbon, the maximum force is about 3.4 mN. For a typical nanoribbon (100 nm in thickness and 1 μm in width), the estimated covalent bonding force is about 1-2 mN, which is comparable order to the maximum tip force (see supporting information). In other words, for nanostructures of small cross-sectional area, the AFM tip can apply enough force to break them. We note that the vertical tip force (~ 10 nN) does not contribute to the exfoliation of $Bi_2Se_3$ nanoribbons – in fact, this force is usually kept small to minimize surface friction and avoid potential damage to the surface of ultra-thin nanoribbon.

The electrical properties of the exfoliated $Bi_2Se_3$ nanoribbons are characterized by both transport and scanning microwave impedance microscopy (MIM) studies. While transport measurement provides quantitative information of individual nanoribbons by lithographically patterning DC electrodes, MIM is much more efficient in obtaining semi-quantitative local conductivity on a large number of pristine materials.[30,31] The setup is implemented on a commercial AFM so that the topographic and electrical information can be obtained simultaneously. In a MIM measurement (Figure 4a), the real (MIM-R) and imaginary (MIM-C) parts of the local sample impedance at 1GHz are detected to form near-field microwave images. The MIM signals can be understood by modeling the impedance response as a function of the sheet resistance ($R_s$) of the nanoribbon (Figure 4b). For a highly conducting ribbon with $R_s < 10$ kΩ/square, a strong MIM-C signal and zero MIM-R signal are expected, taking the Si/SiO$_2$ substrate as a reference. For a highly resistive ribbon with $R_s > 10$ MΩ/square, on the other hand, both MIM-C and MIM-R signals are negligibly small.

Exfoliated Bi$_2$Se$_3$ nanoribbons of five different thicknesses (1, 2, 4, 5, 6 QLs) are scanned by MIM at room temperature. Figure 4c - 4f show the MIM and AFM images and line cuts from a particular Bi$_2$Se$_3$ nanoribbon with different thicknesses. The number of layers (1, 4, 5 QLs) can be identified in the topographic image and the corresponding line cut (Figure 4c, 4e). The simultaneously taken MIM-C image in Figure 4d shows qualitative difference from the AFM data. On one hand, the part of the nanoribbon with 4 and 5 QLs show strong and uniform MIM-C signal and zero MIM-R (line cut, Figure 4f) response, indicating low $R_s < 10$ kΩ/square, consistent with the transport data on the same ribbon (will be discussed later). On the other hand, the part of the nanoribbon with only 1 QL shows little signal in both MIM-C and MIM-R channels, implying very high local $R_s > 10$ MΩ/square. Indeed, we have studied multiple Bi$_2$Se$_3$ nanoribbons with 1 or 2 QLs and measured negligible conductivity by MIM.

Recent theoretical calculations and ARPES measurements have addressed the 3D to 2D crossover of ultra-thin topological insulators[17,18]. For Bi$_2$Se$_3$ down to 1 and 2 QLs, large energy gaps open up in both the bulk and the surface states, whereas for 3QLs and above, the surface gap starts to diminish. Therefore, if the chemical potential happens to fall inside the gap for our 1 and 2 QLs of Bi$_2$Se$_3$, $R_s$ could be quite large even at the room temperature, which agrees with the MIM results. However, there exist other practical issues responsible for the observed high $R_s$. One possibility may be that the 1 or 2 QL nanoribbons are significantly oxidized in the ambient condition. From our experiences, thin Bi$_2$Se$_3$ nanoribbons appear very sensitive to moisture and oxygen in the surrounding environment. Further experiments are needed to distinguish the causes for the observed insulating behavior of 1 and 2 QLs of Bi$_2$Se$_3$. The low $R_s$ in the 4-6 QLs samples can be explained by the small surface gaps and the selenium (Se) vacancies, which work as effective n-type dopants in Bi$_2$Se$_3$ crystals. It is likely that the chemical potential is actually located in the bulk conduction band.[11,13] For the 4-6 QLs, even if the outmost 1-2 QLs are affected by the environment and become insulating, there remain enough Bi$_2$Se$_3$ QLs in the middle that contribute to the small $R_s$.

DC transport measurements are also performed on ultra-thin nanoribbons. A device with 4-probe geometry is fabricated from an AFM-exfoliated $Bi_2Se_3$ nanoribbon with 5 QLs by standard electron beam lithography and thermal evaporation of Cr/Au contacts (Figure 5b). At room temperature, the measured $R_s$ (2.8 k$\Omega$/square) is consistent with the MIM results in the previous section. The transport data of another device fabricated on an as-grown nanoribbon (82 nm thick) are also shown in Figure 5a for comparison. We note that the room temperature conductivity ($\sigma=1/R_s t$) difference between this thick ribbon (3.7 x $10^4$ S/m) and the exfoliated device described above (7.2 x $10^4$ S/m) falls well within the variation from same batch of VLS growth.

Interestingly, these two devices behave differently in the temperature dependence. The resistance of as-grown nanoribbon decreases as temperature goes down to 2 K ($R_s$ = 127 $\Omega$/square), which is typically observed in most of $Bi_2Se_3$ nanoribbon devices (Figure 5c, 5d)[14]. This metallic behavior of $Bi_2Se_3$ nanoribbons is expected for heavily doped semiconductors as discussed earlier. On the other hand, the resistance of the 5 QLs $Bi_2Se_3$ nanoribbon increases as temperature goes down. Unlike the usual band insulator, however, its resistance does not increase exponentially at low temperature, but saturates toward a finite number $R_s$ = 4.9 k$\Omega$/square at our lowest temperature of 2K (Figure 5c, 5d). Such non-metallic behavior is not reported in exfoliated $Bi_2Se_3$ before. Since the outmost 1-2 QLs are shown to be insulating due to surface chemistry in the ambient environment, it is possible that the chemical potential in the remaining intact layers gets close to bulk and surface gaps. The residual conductance is then due to the edge conducting channels only. Further experiments ruling out trivial extrinsic effects are needed to support this hypothesis.

In summary, for the first time, we demonstrate the mechanical exfoliation of layered nanoribbons by AFM, which is very effective and powerful technique to generate ultra-thin layered structure. In the example of $Bi_2Se_3$ nanoribbons, exfoliated layers from 1 QL (~1nm) to multiple QLs are acquired and confirmed by AFM scans on multiple samples. Microwave impedance microscopy reveals the local electronic property of pristine $Bi_2Se_3$ layers and shows that exfoliated nanoribbons thicker than 3 QLs are highly conductive, which is consistent with the room temperature transport measurement. Non-metallic temperature dependence of resistance, distinct from that of as-grown $Bi_2Se_3$ nanoribbons, is observed in the ultra-thin $Bi_2Se_3$ device, leaving an interesting subject for future study.

**Acknowledgement**

Y. C. acknowledges the support from the Keck Foundation. This work is also made possible by the King

Abdullah University of Science and Technology (KAUST) Investigator Award (No. KUS-l1-001-12) and KAUST GRP Fellowship (No. KUS-F1-033-02), NSF Grant DMR-0906027, and Center of Probing the Nanoscale, Stanford University (NSF Grant PHY-0425897).

**Supporting Information Available:** Sample preparation method, characterization methods, and Microwave / transport measurement details. Additional force calculations and data. This material is available free of charge via the Internet at http://pubs.acs.org.

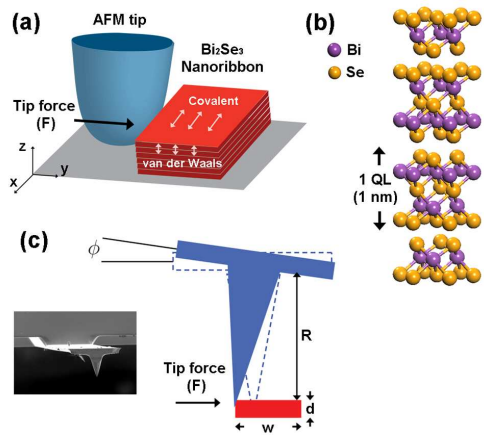

Figure 1. (a) Schematic of AFM exfoliation of layered structure nanomaterial – $Bi_2Se_3$ nanoribbon. Horizontal tip force (y-direction) is applied on the side of a nanoribbon to break the in-plane covalent bonding. (b) Crystal structure of $Bi_2Se_3$. 1 quintuple layer (QL) consists of 5 atomic layers of Bi and Se (Se-Bi-Se-Bi-Se). (c) The horizontal tip force is from torsional displacement ($\varphi$) of the tip. When the tip is scanning across a nanoribbon without z-directional feedback, the whole cantilever is twisted so that the tip can go over step edge $d$. The shear stress on the AFM cantilever generates a torque on the tip end, which applies the horizontal tip force to the nanoribbon.

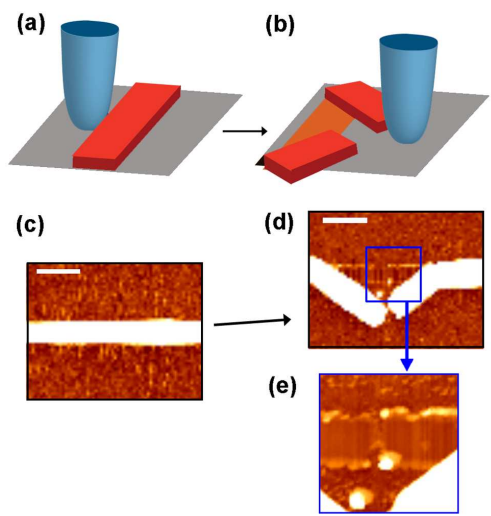

Figure 2. (a), (b) While the tip force breaks covalent bonding of upper layers, weak interlayer bonding - van der Waals bonding is also broken in the middle of the nanoribbon. As a result, several bottom layers are remained intact while the AFM tip breaks most of the Bi$_2$Se$_3$ layers. (c)-(e) AFM topographic image of a Bi$_2$Se$_3$ nanoribbon (d ~ 50 nm) before breaking (c) and after the breaking (d), and the zoom-in image (e). The bottom layers are not broken apart and maintain the same thickness (~2 nm). All scale bars indicate 500 nm.

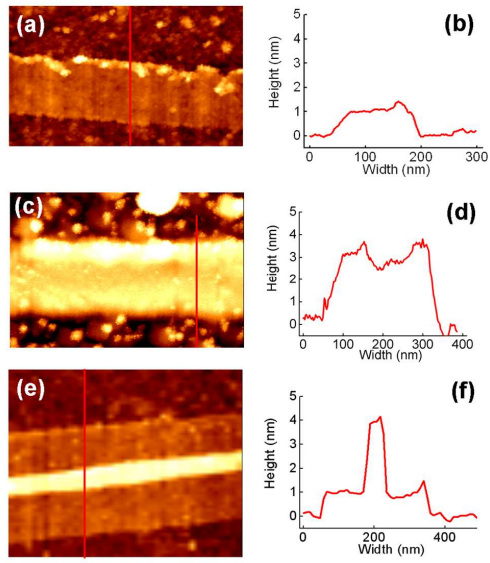

Figure 3. Topographic images and height profiles of very thin Bi$_2$Se$_3$ nanoribbons exfoliated by AFM, corresponding to 1 QL - (a), (b), 3 QLs - (c), (d), and 1 QL/4 QLs - (e), (f). Red lines indicate the location of the height profiles.

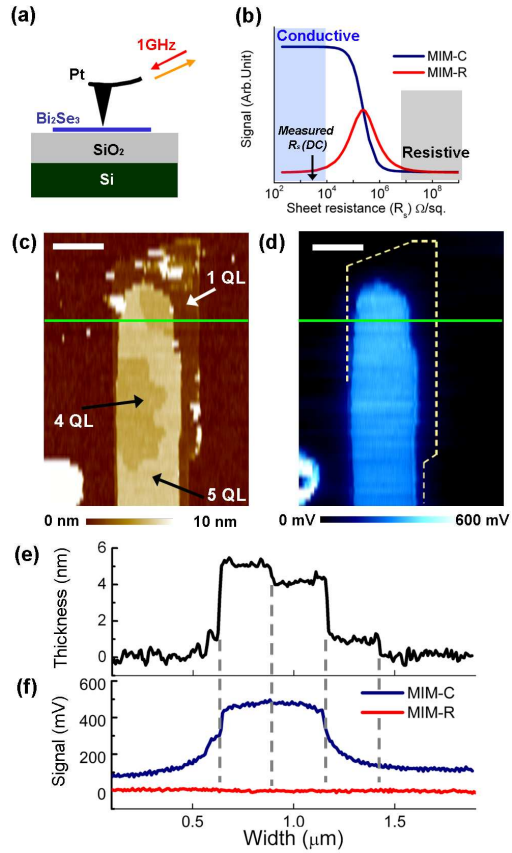

Figure 4. (a) Schematic of the microwave impedance microscopy. The $Bi_2Se_3$ sample is on Si/300 nm $SiO_2$ substrate. A Pt tip scans on the sample surface and a microwave signal (f = 1 GHz) is guided to the tip apex. The impedance responses are recorded as capacitive signal (MIM-C) and resistive signal (MIM-R) after phase-sensitive demodulation. (b) Simulated MIM signals with respect to the Si/ $SiO_2$ substrate saturates for both low ($R_s$ < 10 kΩ/square) and high $R_s$ > 10 MΩ/square) sheet resistance and decreases monotonically in between, in which MIM-R signal is detected. The measured $R_s$ is from transport data of the same device. (c) Topographic image of an exfoliated nanoribbon with three different thicknesses (1 QL, 4 QLs, 5 QLs) in different regions. (d) MIM-C image of the nanoribbon in (c). Strong and uniform MIM-C signals are seen in the 4 and 5 QLs regions, consistent with the low resistivity. On the other hand, 1 QL (edge of the ribbon indicated by broken line) area shows little MIM-C signal, implying high local resistivity. Scale bars in (c), (d) are 500 nm and green lines correspond to the profiles in (e), (f). (e) A line cut from (c), showing 1, 4, and 5 QLs on the substrate. (f) Microwave signals at the same line. Note that the MIM-C signal on the 1 QL section is caused by parasitic capacitance form the nearby metallic 4, 5 QLs. The negligible MIM-R signal throughout the ribbon confirms the high conductivity in 4, 5 QLs and low conductivity in 1 QL $Bi_2Se_3$.

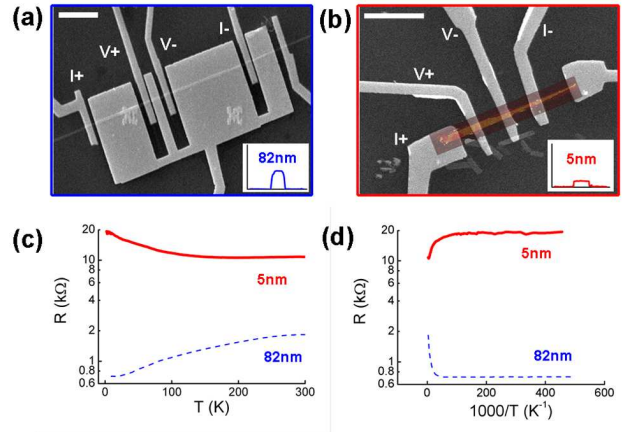

Figure 5. Electrical transport experiments on exfoliated nanoribbon and as-grown nanoribbon. (a) Scanning electron microscopy (SEM) image of the as-grown nanoribbon device with multiple contacts and a front gate (not discussed here). Inset indicates the nanoribbon thickness (82 nm). (b) SEM image of the exfoliated nanoribbon device. Since very thin $Bi_2Se_3$ is barely distinguishable in SEM, AFM image is overlaid to show the device geometry. Several broken segment of the original nanoribbon from exfoliation are seen near the electrodes. A line cut in the inset shows the thickness of this nanoribbon. Both scale bars in (a), (b) are 5um. (c) Resistance R (4-point, log scale) vs. temperature T of the two devices in (a) and (b). R deceases as decreasing T for the as-grown nanoribbon (blue), but increases for the exfoliated nanoribbon (red). (d) R (log scale) vs. 1000/T. The as-grown nanoribbon shows metallic temperature dependence, while the exfoliated nanoribbon's resistance increases during cool-down and saturates near 2K.